\documentclass[12pt]{article}
\pdfoutput=1
\usepackage[nosort]{cite}

\usepackage{draft} 
\usepackage{hyperref}
\usepackage{graphicx,color,subfig,upgreek,mathtools}
\usepackage{amsfonts,amssymb,amsmath,multirow}
\usepackage{mathabx,epsfig}
\usepackage{pifont}
\usepackage{cite}
\usepackage{mciteplus}
\usepackage{skak}
\DeclareFontFamily{OT1}{pzc}{}
\DeclareFontShape{OT1}{pzc}{m}{it}{<-> s * [1.10] pzcmi7t}{}
\DeclareMathAlphabet{\mathpzc}{OT1}{pzc}{m}{it}
 \usepackage{slashed}
\usepackage{xcolor}
\usepackage{environ}

\usepackage{tikz-cd}
\usepackage{datetime}
\NewEnviron{eqs}{%
\begin{equation}\begin{split}
    \BODY
\end{split}\end{equation}
}

\graphicspath{{./Figures/}}

\def\be#1\ee{\begin{align}#1\end{align}}

\def\be{\boldsymbol e}

\def\tilde{\widetilde}
\renewcommand{\bar}{\overline}
\renewcommand{\hat}{\widehat}
\def\^{{\wedge}}
\def\*{{\star}}

\definecolor{ao}{rgb}{0.13, 0.55, 0.13}

\usepackage[normalem]{ulem} 

\definecolor{kspink}{RGB}{200,0,200}

\definecolor{arpit}{RGB}{127,0,0}

\begin{document}

\begin{titlepage}

\begin{center}

\hfill \\
\hfill \\
\vskip 1cm

\title{Time-Reversal Anomalies of QCD$_3$ and QED$_3$}

\author{
Po-Shen Hsin${}^{1,2}$
}

\address{${}^1$ Mani L. Bhaumik Institute for Theoretical Physics,
475 Portola Plaza, Los Angeles, CA 90095, USA}

\address{${}^2$ Department of Mathematics, King’s College London, Strand, London WC2R 2LS, UK.}

\end{center}

%


\abstract{
Anomalies of global symmetry provide powerful tool to constrain the dynamics of quantum systems, such as anomaly matching in the renormalization group flow and obstruction to symmetric mass generation. 
In this note we compute the anomalies in 2+1d time-reversal symmetric gauge theories with massless fermions in the fundamental and rank-two tensor representations, where the gauge groups are $SU(N),SO(N),Sp(N),U(1)$.
The fermion parity is part of the gauge group and the theories are bosonic. 
The time-reversal symmetry satisfies $T^2=1$ or $T^2={\cal M}$ where ${\cal M}$ is an internal magnetic symmetry.
 We show that some of the bosonic gauge theories have time-reversal anomaly with $c_-\neq 0$ mod 8 that is absent in fermionic systems. 
The anomalies of the gauge theories can be nontrivial even when the number of Majorana fermions is a multiple of 16 and $\nu=0$.
}

\vfill

\today

\vfill

\end{titlepage}

\eject

\setcounter{tocdepth}{3}
 \tableofcontents
\bigskip

\section{Introduction}

Gauge theories in 2+1d have broad applications such as deconfined quantum criticalities \cite{Senthil_2004,Wang:2017txt} and anyon theories. A powerful tool to study their dynamics is symmetries and 't Hooft anomalies. For instance, the presence of anomalies forbids  symmetric mass generation \cite{PhysRevX.8.011026,Tong:2021phe,Wang:2022ucy} that makes the theory flow to a trivially gapped phase while preserving the symmetry.

In this note we will study the anomalies in time-reversal symmetric gauge theories with massless charged fermions. We will focus on the theories that are bosonic, i.e. all the local operators have integer spins. In such theories the fermion parity symmetry $(-1)^F$ is gauged and belongs to part of the gauge group. We will call such gauge theories a bosonic gauge theory with charged fermions.
In addition to the time-reversal symmetry, the theories can have other global symmetry such as flavor symmetry and magnetic symmetry. 

The anomalies of time-reversal symmetry in gauge theories with charged fermions have been studied extensively in the literature (e.g. \cite{Gomis:2017ixy,Cordova:2017kue,Wang:2017txt,PhysRevX.10.011033,Calvera:2021jqi,Zou:2021dwv,Delmastro:2022pfo}). However, there were several open problems:
\begin{itemize}
    \item[1. ] The method of computing anomalies often counts the anomalies of free fermions with $T^2=(-1)^F$ symmetry. 
    Namely, one can start with free fermions and turn on the gauge fields.
    However, as the theories are bosonic, there can be anomalies that are not visible in fermionic systems. We will review such  missing bosonic anomalies in section \ref{sec:reviewbosonicTanomaly}.

    \item[2. ] Related to the above issue, there can be contribution to the anomalies from the gauge fields in addition to the anomalies from the fermions. In fact, even pure gauge theories without charged fermions can have nontrivial anomalies, as described by anomalies in topological orders (e.g. \cite{Wang:2016qkb,Barkeshli:2016mew,Tachikawa:2016cha,Tachikawa:2016nmo}). 

    \item[3. ] In gauge theories with monopoles, the time-reversal symmetry can be deformed to be an extension, such as $T^2={\cal M}(-1)^F$ for $\mathbb{Z}_2$ subgroup magnetic symmetry ${\cal M}$ (see e.g. \cite{Cordova:2017kue}). In such cases, the time-reversal anomalies have not been computed in gauge theories with charged massless fermions. In particular, the classification of anomalies for such symmetry is only understood recently \cite{Cordova:2023bja,Debray:2023iwf}, and the anomaly is computed for simple TQFTs \cite{Cordova:2023bja,Debray:2023iwf}.
    
\end{itemize}

In this note we will demonstrate a method of computing the anomalies that address these issues, using gauge theories with $SU(N),Sp(N),U(1),SO(N)$ gauge groups and fermions in the fundamental and rank-two representations. The symmetries discussed here are invertible symmetries. In a upcoming paper we will discuss non-invertible symmetries. There are various proposed dynamics of these gauge theories, many of them involve Goldstone bosons, e.g. \cite{Komargodski:2017keh,Choi:2018tuh}. It will be interesting to constrain the dynamics using anomalies in future work.

\subsection{Anomalies of $T^2=1$ in 2+1d bosonic theories}
\label{sec:reviewbosonicTanomaly}

In many examples of time-reversal symmetric bosonic gauge theories with charged fermions, the time-reversal symmetry $T$ satisfies $T^2=1$, since $(-1)^F=1$ is absent. The anomalies of such time-reversal symmetry admit $\mathbb{Z}_2\times \mathbb{Z}_2$ classification, described by the 3+1d bulk SPT phase with effective action \cite{Kapustin:2014tfa}
\begin{equation}
    \alpha \pi\int w_2^2+\beta \pi\int w_1^4~,
\end{equation}
where $\alpha,\beta=0,1$, and $w_i$ are the $i$th Stiefel-Whitney class of the tangent bundle. 
The anomaly $\alpha$ can be detected by $\mathbb{CP}^2$ using $\int_{\mathbb{CP}^2} w_2^2=1,\int_{\mathbb{CP}^2} w_1^4=0$ mod 2, while the anomaly $\beta$ can be detected by $\mathbb{RP}^4$ using $\int_{\mathbb{RP}^4} w_2^2=0,\int_{\mathbb{RP}^4} w_1^4=1$ mod 2.
Our goal is to determine $\alpha,\beta$ for various bosonic 2+1d gauge theories with charged fermions.

The first coefficient $\alpha$ captures the chiral central charge $c_-=4\alpha$ mod 8, which can be defined by the unambiguous fractional part in the stress tensor two-point function that is not plagued by the $c\in 8\mathbb{Z}$ counterterm: \cite{Closset:2012vp}
\begin{equation}
\langle T_{\mu\nu}(x)T_{\rho\sigma} (0)\rangle    \supset \frac{ic_-}{384\pi}\left(\left(\epsilon_{\mu\rho\lambda}\partial^\lambda(\partial_\nu\partial_\sigma-\eta_{\nu\sigma}\partial^2)+(\mu\leftrightarrow\nu)\right)+(\rho\leftrightarrow \sigma)\right)\delta^3(x)~,
\end{equation}
where the above correlation function is on Euclidean signature spacetime.
To see this, we note that the contact term is parity odd, therefore the correlation function violates the time-reversal symmetry. 
If we want to preserve the time-reversal symmetry by cancelling the contact term, we would need additional fractional gravitational Chern-Simons term $\text{CS}_\text{grav}$ with coefficient $2c_-$. However, such gravitational Chern-Simons term with $2c_-\neq 0$ mod 16 is not well-defined, and it depends on the bulk by
\begin{equation}
    -\frac{2\pi c_-}{8}\int (p_1/3),\quad p_1=-\frac{1}{2(2\pi)^2}\text{Tr}(R\wedge R)~,
\end{equation}
where $R$ is the curvature 2-form.
When $c_-=0$ mod 4, this is $\alpha \pi \int w_2^2$ with $\alpha=c_-/4$ mod 2.

The relation between anomaly $\alpha$ and chiral central charge is previously obtained for 2+1d topological orders, see. e.g. \cite{Wang:2016qkb,Barkeshli:2016mew,Tachikawa:2016cha,Tachikawa:2016nmo}).

\subsubsection{Reduction of anomalies in fermionic systems}

If we regard the bosonic system as a decoupled part of a fermionic system, the anomalies can become trivial.
The anomaly $\alpha$ is absent in fermionic systems with $T^2=(-1)^F$, where the suitable background has $w_2=0$ \cite{Kapustin:2014dxa}. Similarly, $\alpha,\beta$ are not independent in fermionic systems with $T^2=+1$, where the suitable background has $w_2=w_1^2$ \cite{Kapustin:2014dxa}.
In particular, the anomalies $\alpha$ cannot be obtained from free fermions with $T^2=(-1)^F$.

The second anomaly coefficient $\beta$ represents anomalies that can also appear in free fermion systems with $T^2=(-1)^F$, where the anomalies admit $\nu\in \mathbb{Z}_{16}$ classification \cite{Fidowski:2013,Wang_2014,Metlitski:2014,Kapustin:2014dxa,kitaev:2015lecture}, and the bosonic anomaly $\beta$ corresponds to the class $\nu=8\beta$ mod 16.

\subsection{Computation of anomaly in gauge theories with charged fermions}

The partition function of massless fermions coupled to gauge field $a$ has a phase in the regularization of \cite{Witten:2015aba}:
\begin{equation}
Z=|Z|e^{-\pi i\eta(a)/2}    ~,
\end{equation}
where $\eta(a)$ is the eta invariant of the covariant Dirac operator with gauge field $a$.
When there are even number of fermions, the phase can be expressed as a Chern-Simons term using Atiyah-Patodi-Singer index theorem \cite{atiyah_patodi_singer_1975}
\begin{equation}
    e^{-\pi i\eta(a)} =\exp\left(2\pi i \text{Tr}\int_Y e^{-F/2\pi}{\hat A}(R)\right)~,
\end{equation}
where $Y$ is a four-manifold whose boundary is the 2+1d spacetime, $F$ is the field strength and $R$ is the Ricci 2-form.

For example, the partition function of two massless Dirac fermions coupled to the same $U(1)$ gauge field $A$ in this regularization has the phase
\begin{align}
    2\pi\int_Y e^{F/2\pi} {\hat A}(R)
    &=2\pi\int_Y \left(
    \frac{1}{2!}\left(\frac{F}{2\pi}\right)^2
+{\hat A}(R)
    \right)
=\int_Y\left(
\frac{1}{4\pi}dAdA
\right)+\frac{2\pi\sigma}{8}\cr 
&=\int_Y\left(
\frac{1}{4\pi}dAdA
+\frac{1}{96\pi}\text{Tr}\left(R\wedge R\right)\right)~,
\end{align}
where $\sigma$ is the signature. We note that the contribution of $\sigma$ comes with a positive sign. Here, the convention for signature is such that $\sigma=-p_1/3$ on four-manifolds.

We will use time-reversal symmetry $T$ that commutes with the gauge group, i.e. the electric charge of the gauge field is odd under time-reversal. By combining $T$ with another internal unitary symmetry $U$ we can obtain another anti-unitary symmetry $T'=TU$ which may or may not commute with the gauge group. The anomaly of $T'$ is completely specified by the anomaly of the time-reversal symmetry $T$ and the internal symmetry, thus we will focus on symmetry $T$ without loss of generality.

To preserve time-reversal symmetry, we should cancel the phase of the partition function with additional Chern-Simons counterterm term. 
 When the gauge bundle is twisted by a quotient on the gauge group, the Chern-Simons term is not well-defined and depends on the bulk, which gives the anomaly of the theory.
The theories we consider have the symmetry structure
\begin{equation}
    {G_\text{gauge}\times \tilde G_\text{global}\over C}~,
\end{equation}
where $C$ is in the center, $G_\text{gauge}$ is the gauge group, and the global symmetry is $\tilde G_\text{global}/C=G_\text{global}$ which includes both the internal symmetry and the Lorentz group. The quotient means that in the presence of $G_\text{global}$ background gauge field that is not a $\tilde G_\text{global}$ background gauge field, the background gauge fields are twisted and they have fractional instanton number. As a consequence, in the presence of such background gauge field, the Chern-Simons counterterm that we use to cancel the phase of the partition function is not well-defined but depends on the bulk.
Examples of such method for computing anomalies are discussed in e.g. \cite{Benini:2017dus,Cheng:2022nji}.
The anomalies include the contribution from both the fermions and the gauge field. 

If the gauge field has additional time-reversal symmetric topological terms, then there can be additional contribution to the anomalies from these terms. These contributions are important for anomalies in topological orders described by pure gauge theories, but we will not
 discuss such examples here unless in gauge theories with magnetic symmetry, where the coupling to the background gauge field of the magnetic symmetry is a topological term.

\subsection{Summary of results}

Let us list the time-reversal anomalies we computed in bosonic gauge theories with charged fermions. Here, we list the anomaly for time-reversal symmetry $T$ that commutes with the gauge group and the flavor symmetry. The anomaly for any anti-unitary symmetry $T'$ can be obtained from the anomaly of $T$ and internal symmetry.
\begin{itemize}
    \item $G=SU(N)$ with $N_f$ fundamental fermions. $N,N_f$ need to be even. The anomalies for $T^2=1$ are $\alpha=0,\beta=NN_f/4$.
    The anomaly including the flavor symmetry is in (\ref{eqn:fullanomsu}).
    
    \item $G=Sp(N)$ with $N_f$ fundamental fermions. $N_f$ needs to be even.
    The anomalies for $T^2=1$ are $\alpha=0,\beta=NN_f/2$.
    The full anomaly including the flavor symmetry is in (\ref{eqn:fullanomsp}) for odd $N$. When $N$ is even, the theory does not have anomaly.
    
    \item $G=SO(N)$ with $N_f$ vector fermions. $N_f$ needs to be even. 
 \begin{enumerate}
     \item When $N_f=2$ mod 4, the time-reversal symmetry satisfies $T^2={\cal M}$ for $\mathbb{Z}_2$ (subgroup) magnetic symmetry ${\cal M}$ \cite{Cordova:2017kue}. 
     The time-reversal symmetry by itself is non-anomalous, i.e. trivial class in the
     $\mathbb{Z}_4\times \mathbb{Z}_4$ classification \cite{Cordova:2023bja}.
     The full anomaly including the flavor symmetry is in (\ref{eqn:fullanomso2mod4}).

     \item When $N_f=2$ mod 4, the time-reversal symmetry satisfies $T^2=1$ \cite{Cordova:2017kue}. The time-reversal anomaly is $\alpha=0$ and $\beta=NN_f/8$. The full anomaly including the flavor and magnetic symmetry is in (\ref{eqn:fullanomso0mod4}).

 \end{enumerate}

    \item $G=U(1)$ with $N_f$ charge-one fermions. $N_f$ needs to be even. When $N_f=2$ mod 4, the time-reversal symmetry satisfies $T^2=(-1)^m$ for magnetic charge $m$. The time-reversal symmetry is non-anomalous from
    the classification of \cite{Cordova:2023bja}. when $N_f=0$ mod 4, the time-reversal symmetry satisfies $T^2=1$. The time-reversal anomaly is $\alpha=0,\beta=N_f/4$.

\item $G=SU(16n+8)$ with a Dirac fermion in the 2-index symmetric or antisymmetric tensor representation. The time-reversal anomaly is $\alpha=1,\beta=1$.

\item By stacking $SU(16n+8)$ gauge theory with rank-2 representation and e.g. $SU(2)$ gauge theory with two fundamental fermions, we obtain bosonic theories with $\alpha=1,\beta=0$. The combined bosonic theory has a multiple of 16 Majorana fermions and $\beta=0$, but the time-reversal anomaly $w_2^2$ is nontrivial $\alpha=1$. 
The theories have $c_-=4$ mod 8 anomaly, which can be different from free massless fermions contribution.
\end{itemize}

The note is organized as a set of examples. In section \ref{sec:SU}-\ref{sec:SO} we will discuss the time-reversal anomalies in bosonic gauge theories with charged fermions with $G=SU(N),Sp(N),SO(N)$ gauge groups, respectively. In section \ref{sec:U(1)} we will discuss time-reversal anomalies in QED$_3$, which is also a bosonic theory. 
In section \ref{sec:ranktwo} we discuss gauge theories with fermions in rank-two representation and nontrivial $w_2^2$ anomaly.
In Appendix \ref{sec:mathidentity} we summarize some useful identities for characteristic classes.

\section{$SU(N)$ gauge theory with $N_f$ fundamental fermions}
\label{sec:SU}

Let us consider time-reversal symmetric QCD$_3$ with $SU(N)$ gauge group and $N_f$ massless fermions in the fundamental representation. We will focus on even $N,N_f$. 
Such theories are bosonic: the fermion parity $(-1)^F$ that flips the sign of all fermions is gauged. All gauge-invariant local operators are bosons.

\subsection{Symmetry}

The theory is bosonic: since $N$ is even, the fermion parity is part of the gauge group.
The symmetry structure is:
\begin{equation}
    {SU(N)_\text{gauge}\times U(N_f)_\text{flavor}\times \widetilde{\text{Lorentz}}\over \mathbb{Z}_N\times \mathbb{Z}_2}~,
\end{equation}
where the continuous flavor symmetry includes the $U(1)$ baryon number, and $\widetilde{\text{Lorentz}}$ is an extension of the Lorentz group depending on how time-reversal fractionalizes on the charged fermion, i.e. $T^2=(-1)^F$ or $T^2=+1$. We will take the fractionalization $T^2=(-1)^F$ in the following discussion. The above time-reversal symmetry commutes with all the other symmetries.
The global 0-form symmetry is
\begin{equation}
    {U(N_f)_\text{flavor}\times \widetilde{\text{Lorentz}}\over \mathbb{Z}_N\times \mathbb{Z}_2}=U(N_f)/\mathbb{Z}_N\times \text{Lorentz}~.
\end{equation}

The background gauge field for $U(N_f)/\mathbb{Z}_N=\left(SU(N_f)\times U(1)\right)/\left(\mathbb{Z}_{N_f}\times \mathbb{Z}_N\right)$ is described by the following relation between the first Chern class $c_1$ and $\mathbb{Z}_N,\mathbb{Z}_{N_f}$ valued magnetic fluxes $w_2^{f},w_2'^{f}$: \cite{Benini:2017dus}
\begin{equation}\label{eqn:NNfbundle}
    c_1=\frac{N}{\ell} w_2'^{f}+\frac{N_f}{\ell} w_2^{f}\text{ mod }\frac{NN_f}{\ell}~,
\end{equation}
where $\ell=\gcd(N,N_f)$ is even.

The quotient in the symmetry structure implies that  the gauge bundle is twisted to be a $SU(N)/\mathbb{Z}_N$ bundle.
\begin{equation}
    w_2^\text{gauge}=w_2^f+\frac{N}{2}x_2\text{ mod }N~,
\end{equation}
where $x_2$ depends on how the $T^2=1$ time-reversal symmetry fractionalizes on the charged fermions:
$x_2=w_2$ for $T^2=(-1)^F$ and $x_2=w_2+w_1^2$ for $T^2=1$ \cite{Kapustin:2014dxa}. In the discussion we will focus on the fractionalization $T^2=(-1)^F$ where $x_2=w_2$.

\subsection{Anomaly}

To cancel the phase of the partition function, we need to add the counterterm
\begin{equation}
    \frac{SU(N)_{N_f/2}\times U(N_f)_{N/2}\times NN_f\text{CS}_\text{grav}}{\mathbb{Z}_N\times \mathbb{Z}_2}~,
\end{equation}
where $\text{CS}_\text{grav}$ is the gravtiational Chern-Simons term $\int_{\partial Y} \text{CS}_\text{grav}=-\frac{\pi}{8}\int_Y (p_1/3)$.
The above Chern-Simons term is not well-defined in the presence of discrete fluxes $w_2^f,w_2'^f$. The bulk dependence is
\begin{align}\label{eqn:fullanomsu}
    &2\pi\frac{N_f}{2}\frac{N-1}{2N}\int {\cal P}(w_2^f+\frac{N}{2}w_2)+2\pi \frac{N}{2}\frac{N_f-1}{2N_f}\int {\cal P}(w_2'^f)+\frac{NN_f \pi/2}{(NN_f/\ell)^2} \int c_1^2-\frac{NN_f\pi}{8}\int (p_1/3)\cr 
    &=2\pi \frac{N_f}{2}\frac{N-1}{2N}\int {\cal P}(w_2^f)
    +\frac{N_f}{2}\pi\int w_2^f\cup w_2+2\pi \frac{N}{2}\frac{N_f-1}{2N_f}\int {\cal P}(w_2'^f)\cr
    &\quad + \frac{\pi\ell/2}{NN_f/\ell}\int \left(\frac{N^2}{\ell^2}{\cal P}(w_2'^f)+\frac{N_f^2}{\ell^2}{\cal P}(w_2^f)+\frac{2NN_f}{\ell^2}w_2^f\cup w_2'^f\right)
    \cr
    &\quad 
    +\frac{NN_f(N-2)}{8}\pi\int w_2^2
    -\frac{NN_f}{4}\pi\int w_1^4
    \cr
    &=\frac{N_f}{2}\pi\int (w_2^f)^2+\frac{N_f}{2}\pi\int w_2^f\cup w_2+\pi\int w_2^f\cup w_2'^f+\frac{N}{2}\pi\int (w_2'^f)^2+\frac{NN_f}{4}\pi\int w_2^2+\frac{NN_f}{4}\pi\int w_1^4\cr 
    &=\frac{N_f}{2}\pi\int w_2^f\cup w_1^2+\pi\int w_2^f\cup w_2'^f+\frac{N}{2}\pi\int w_2'^f\cup (w_2+w_1^2)+
    \frac{NN_f}{4}\pi\int w_1^4~,
\end{align}
where in the second line we used $p_1=3{\cal P}(w_2)+2w_1^4$ mod 4 (see Appendix \ref{sec:mathidentity}). In the third line we used 
\begin{equation}
    \frac{NN_f(N-2)}{8}=\frac{N_f}{2}\frac{N}{2}\left(\frac{N}{2}-1\right)\in 2\mathbb{Z}~.
\end{equation}
In the last line we used the Wu formula $x_2\cup x_2=x_w\cup (w_2+w_1^2)$ mod 2.

The anomaly of time-reversal symmetry alone is the last term in (\ref{eqn:fullanomsu}), which corresponds to $c_-=0$ mod 8 and $\nu=2NN_f\in 8\mathbb{Z}$. We note that while $\nu$ agrees with the massless free fermion answer, $c_-$ is different from the massless free fermion answer $c^\text{free}_-=\frac{1}{2}\times NN_f=NN_f/2$.
The first term is a mixed anomaly between time-reversal symmetry and flavor symmetry that can be detected on unorientable manifolds $w_1\neq 0$ of the type discussed in \cite{Witten:2016cio}.

We note that when $NN_f=0$ mod 8, $\nu=0$. On the other hand, there can still be anomalies involving the flavor symmetry. Thus the time-reversal and flavor symmetry is an obstruction to symmetric mass generation without breaking the flavor symmetry even when the system has a multiple of 16 Majorana fermions.

\subsection{Examples}

\subsubsection{Example of mixed anomaly: $SU(2)$ QCD with $N_f=2$}

For instance, $SU(2)$ QCD$_3$ with $N_f=2$ flavors has the anomaly\footnote{
We note that the $\pi\int (w_2^f)^2$ term happens to be trivial for $N_f=2$ by dimension reason: the cohomology classes are $H^*(BSO(3),\mathbb{Z}_2)=\mathbb{Z}_2[w_2,w_3]$ (see e.g. \cite{milnor1974characteristic,Brown1982}).
}
\begin{equation}
    \pi\int \left(w_1^4+w_2^f\cup w_1^2+w_2^f\cup w_2'^f+(w_2'^f)^2\right)~.
\end{equation}
If we break the flavor symmetry to the $SU(2)/\mathbb{Z}_2=SO(3)$ subgroup, $w_2^f=w_2'^f$, and the anomaly becomes
\begin{equation}
    SO(3)\text{ flavor symmetry:}\quad 
    \pi\int \left(w_1^4+w_2^f\cup w_1^2\right)~.
\end{equation}
This anomaly with the $SO(3)$ flavor symmetry was discussed in \cite{Wang:2017txt}, although the decomposition into the $w_1^4$ and $w_2^2$ parts are not discussed there.

Note that if we gauge the $SO(3)$ flavor symmetry, the gauge group becomes $SO(4)$ with one Dirac fermion (i.e. $N_f'=2$ Majorana fermions) in the vector representation. As we will discuss in section \ref{sec:SO}, the mixed anomaly
\begin{equation}
    \pi\int w_2^f\cup w_1^2
\end{equation}
implies that the time-reversal symmetry is deformed to the extension $T^2={\cal M}$ where ${\cal M}$ is the $\mathbb{Z}_2$ magnetic symmetry. This agrees with the time-reversal symmetry in $SO(4)$ gauge theory with $N_f'=2$ vector fermions \cite{Cordova:2017kue}. The deformed time-reversal symmetry implies $w_1^2$ is trivial, thus the anomaly is
\begin{equation}
    \pi\int w_2^2~.
\end{equation}
This is the class $\ell=2$ anomaly in the classification of \cite{Cordova:2023bja}, and it agrees with the anomaly in $SO(N)$ gauge theory in (\ref{eqn:fullanomso2mod4}) we will derive in section \ref{sec:SO}.

\subsubsection{Example: $SU(4)$ QCD with $N_f=2$}

Consider time-reversal symmetric $SU(4)$ QCD with $N_f=2$. The continuous $U(2)/\mathbb{Z}_4$ flavor symmetry is anomalous, with anomaly described by
\begin{equation}
    \pi\int w_2^f\cup w_2'^f~,
\end{equation}
We note that time-reversal symmetry by itself is non-anomalous. In particular, $\nu=16=0$ mod 16. Here we show that the anomaly $w_2^2$ is also absent. 

If we break the flavor symmetry to $SU(2)/\mathbb{Z}_2=SO(3)$ symmetry, there is only one independent discrete flux for the background gauge field from the $\mathbb{Z}_4$ valued $w_2^f$ and $\mathbb{Z}_2$ valued $w_2^f$, and the $SU(4)$ gauge bundle can still be twisted to an $SU(4)/\mathbb{Z}_2$ bundle. This corresponds to $w_2'^f=2w_2^f$, and the anomaly becomes trivial.
This theory with $SO(3)$ flavor symmetry has been considered in symmetric mass generation \cite{PhysRevX.8.011026}. On the other hand, if the flavor symmetry is enlarged to be $U(2)/\mathbb{Z}_4$, then the above anomaly implies that there is not symmetric mass generation.

\section{$Sp(N)$ gauge theory with $N_f$ fundamental fermions}
\label{sec:Sp}

Consider $Sp(N)$ gauge theory with $N_f$ massless fermions in the fundamental $\mathbf{2N}$ dimensional representation. 
The fermion parity is gauged, and the theory is bosonic for any $N,N_f$. 
Time-reversal symmetric theory requires $N_f$ to be even.

\subsection{Symmetry}

The symmetry structure is
\begin{equation}\label{eqn:Spsymmetry}
    {Sp(N)_\text{gauge}\times Sp(N_f)\times \widetilde{\text{Lorentz}}\over \mathbb{Z}_2\times\mathbb{Z}_2}~,
\end{equation}
where the extension $\widetilde{\text{Lorentz}}$ of Lorentz group depends on the fractionalization of time-reversal symmetry on the fermions, and we will take the fractionalizaiton to be $T^2=(-1)^F$. The time-reversal symmetry commutes other symmetries.
The global 0-form symmetry is
\begin{equation}
    {Sp(N_f)\times \widetilde{\text{Lorentz}}\over \mathbb{Z}_2\times\mathbb{Z}_2}=Sp(N)/\mathbb{Z}_2\times \text{Lorentz}~.
\end{equation}
Since there are two $\mathbb{Z}_2$ quotients, the background for the symmetry has two $\mathbb{Z}_2$ discrete fluxes. Denote the $\mathbb{Z}_2$ flux of $Sp(N_f)/\mathbb{Z}_2$ by $w_2^f$. The flux of the Lorentz group  depends on the fractionalization of time-reversal symmetry on the charged fermion. For $T^2=(-1)^F$ it is $w_2$, while for $T^2=+1$ it is $w_2+w_1^2$. We will focus on the fractionalization $T^2=(-1)^F$.

The symmetry structure (\ref{eqn:Spsymmetry}) implies that the discrete flux of $Sp(N)/\mathbb{Z}_2$ is given by
\begin{equation}
    w_2^\text{gauge}=    
    w_2^f+w_2~.
\end{equation}

\subsection{Anomaly}

The regulator for the time-reversal symmetric theory is
\begin{equation}
    \frac{Sp(N)_{N_f/2}\times Sp(N_f)_{N/2}\times 2NN_f\text{CS}_\text{grav}}{\mathbb{Z}_2\times\mathbb{Z}_2}~.
\end{equation}
The regulator is not well-defined, but it depends on the bulk, which gives rise to anomalies.

\subsubsection{Time-reversal anomaly}

Let us ignore the flavor symmetry for the moment. The regulator is
\begin{equation}
    \frac{Sp(N)_{N_f/2}\times 2NN_f\text{CS}_\text{grav}}{\mathbb{Z}_2}~.
\end{equation}
It has the bulk dependence
\begin{equation}
    2\pi \frac{NN_f}{8}\int {\cal P}(w_2)+2NN_f \frac{\pi}{8}\int (p_1/3)=
    \frac{2\pi}{4}N\frac{N_f}{2}\int {\cal P}(w_2)-\frac{2\pi}{4} N \frac{N_f}{2}\int (p_1/3)
    ~,
\end{equation}
which depends on $p_1$ mod 4. Using $p_1=3{\cal P}(w_2)+2w_1^4$, we can simplify it into
\begin{equation}
    \frac{NN_f}{2}\pi\int w_1^4~.
\end{equation}

\subsubsection{Full anomaly}

Let us separate the discussion into even and odd $N$.

\paragraph{Even $N$}

When $N$ is even, the bulk term is
\begin{align}
    2\pi& \frac{NN_f}{8}\int {\cal P}(w_2+w_2^f)
    +
    2\pi \frac{NN_f}{8}\int {\cal P}(w_2^f)
    -2NN_f \frac{\pi}{8}\int (p_1/3)\cr 
    &=
    \left(2\pi \frac{NN_f}{8}\int {\cal P}(w_2)
    -2NN_f \frac{\pi}{8}\int (p_1/3)\right)
    +2\pi \frac{NN_f}{4}\int {\cal P}(w_2^f)=0\text{ mod }2\pi~.
\end{align}
Thus there is no anomaly for even $N$.

\paragraph{Odd $N$}

When $N$ is odd, the bulk term is 
\begin{align}\label{eqn:fullanomsp}
 &I_{\theta_f=N\pi}+ 2\pi \frac{NN_f}{8}\int {\cal P}(w_2+w_2^f)-2NN_f \frac{\pi}{8}\int (p_1/3)\cr
 &= 
I_{\theta_f=N\pi}+    \frac{2\pi}{4}\frac{NN_f}{2}\int {\cal P}(w_2^f)+\frac{NN_f}{2}\pi\int w_2\cup w_2^f
+\frac{NN_f}{2}\pi\int w_1^4~,
\end{align}
where $I_{\theta_f=N\pi}$  is the $\theta_f=N\pi$ is the theta term for the $Sp(N_f)/\mathbb{Z}_2$ bundle. In particular there is a mixed anomaly between the flavor symmetry and time-reversal symmetry.
Note that the combination
\begin{equation}
    I_{\theta_f=N\pi}+\frac{2\pi}{4}\frac{NN_f}{2}\int {\cal P}(w_2^f)~
\end{equation}
is invariant under time-reversal symmetry and has order 2. When $N=1$, this agrees with (\ref{eqn:fullanomsu}) for $SU(2)$ gauge group.

\section{$SO(N)$ gauge theory with $N_f$ fundamental fermions}
\label{sec:SO}

Let us consider time-reversal symmetric QCD$_3$ with $SO(N)$ gauge group and $N_f$ fermions in the vector representation. 
We will consider even $N$: then the fermion parity is gauged, and the theory is bosonic. In addition, for the theory to be time-reversal symmetric we will take even $N_f$ to avoid the standard parity anomaly. 

A new feature of the theory is that the local operators of the theory not only include the gauge-invariant combinations of the fermion operators and field strength, there are also monopole operators due to $\pi_1(SO(N))=\mathbb{Z}_2$ for $N>2$ and $\pi_1(SO(2))=\mathbb{Z}$. The monopole operators transform under the magnetic symmetry. When $N=2$, we will focus on the $\mathbb{Z}_2\subset U(1)$ subgroup magnetic symmetry in this section.

\subsection{Symmetry}

The symmetry of the classical Lagrangian is
\begin{equation}
{ SO(N)_\text{gauge}\times SO(N_f)\times \widetilde{\text{Lorentz}}\over \mathbb{Z}_2\times \mathbb{Z}_2}~,
\end{equation}
where the extension $\widetilde{\text{Lorentz}}$ of Lorentz group depends on the fractionalization of time-reversal symmetry on the fermions, and we will take the fractionalizaiton to be $T^2=(-1)^F$. The time-reversal symmetry commutes other symmetries.

The global symmetry of the Lagrangian is
\begin{equation}
{SO(N_f)\times \widetilde{\text{Lorentz}}\over \mathbb{Z}_2\times \mathbb{Z}_2}~.
\end{equation}
The two $\mathbb{Z}_2$ quotients implies that the background gauge field for the symmetry has two $\mathbb{Z}_2$ discrete fluxes, given by $w_2^f$ for $SO(N_f)/\mathbb{Z}_2$ and $w_2$.
As we will show below, if we include the magnetic 0-form symmetry that transforms the monopole operators, the global symmetry is further extended. 

The true global symmetry includes the $\mathbb{Z}_2$ magnetic symmetry that transforms the monopole of $SO(N)$ gauge theory.
The monopole is dressed with $N_f/2$ fermions with the flavor indices symmetrized. Therefore when $N_f=2$ mod 4, the center of $SO(N_f)$ flavor symmetry acts on the monopoles and there is no $\mathbb{Z}_2$ quotient.\footnote{
Another explanation is ${SO(N_f)\times \mathbb{Z}_2^{\cal M}\over \mathbb{Z}_2}\cong SO(N_f)$, where $\mathbb{Z}_2^{\cal M}$ is the magnetic symmetry.

} In particular, the center is identified with the $\mathbb{Z}_2$ magnetic symmetry.
On the other hand, if $N_f=0$ mod 4, the monopoles also do not transform under the center of $SO(N_f)$, but there is additional $\mathbb{Z}_2$ magnetic symmetry. 
If we denote $B$ to be the background of the magnetic symmetry, for $N_f=2$ mod 4, $w_2^f=dB$, while for $N_f=0$ mod 4 $w_2^f$ is nontrivial with an independent $B$ that satisfies $dB=0$.

\subsection{Anomaly}

To make the theory time-reversal symmetric, we need to add the regulating term
\begin{equation}\label{eqn:regSOSO}
 SO(N)_{N_f/2}\times SO(N_f)_{N/2}\times \frac{NN_f}{2}\text{CS}_\text{grav}~,
\end{equation}
where we omit the quotient that depend on $N_f$ mod 4.
Such term is not well-defined, but depends on the bulk:
\begin{align}
    &2\pi\frac{NN_f/2}{16}\int {\cal P}(w_2+w_2^f)+\frac{N_f}{2}\pi\int \left(w_2(SO(N))\right)^2+\frac{N_f}{2}\pi\int w_2(SO(N)\cup (w_2+w_2^f)\cr 
    &+2\pi \frac{NN_f/2}{16}\int {\cal P}(w_2^f)+\frac{N_f}{2}\pi\int w_2(SO(N_f))\cup w_2^f+\frac{N}{2}\pi\int \left(w_2(SO(N_f)) \right)^2\cr 
    &-\frac{NN_f}{2}\frac{\pi}{8}\int (p_1/3)+S_\text{mag}~,
\end{align}
where the first three lines respectively represent the bulk dependence of (\ref{eqn:regSOSO}), and
$S_\text{mag}$ is the contribution from the $\mathbb{Z}_2$ magnetic symmetry determined as follows. 

The background gauge field $B$ for the magnetic symmetry couples to the theory by the topological term
\begin{equation}
    \pi\int_{3d} w_2(SO(N))\cup B~.
\end{equation}
Since the gauge bundle is twisted, $dw_2(SO(N))=\frac{N}{2}\text{Bock} \left(w_2+w_2^f\right)$ with Bock$=\frac{d}{2}$ being the Bockstein homomorphism for $2\mathbb{Z}\rightarrow\mathbb{Z}\rightarrow\mathbb{Z}_2$. Thus the above topological term is not well-defined, but depends on the bulk by
\begin{equation}
    S_\text{mag}=\frac{N}{2}\pi\int \left(w_3^f+w_3\right)\cup B
\end{equation}
where $B$ is the background gauge field for the magnetic symmetry, and we used $\text{Bock}(w_2)=w_3+w_1\cup w_2=w_3$ since $w_1\cup w_2=0$ on 4-manifolds, and similarly $\text{Bock}(w_2^f)=w_3^f$. The first term is discussed in  \cite{Cordova:2017vab}.

For the bulk term to be independent of the dynamical field $w_2(SO(N))$, the symmetry is the following:
\begin{itemize}
    \item For $N_f=2$ mod 4, the background must satisfies $w_1^2=dB$. This implies that the time-reversal symmetry satisfies
    \begin{equation}
        N_f=2\text{ mod }4:\quad T^2={\cal M}=z~,
    \end{equation}
    where ${\cal M}$ is the magnetic symmetry, and $z$ is the $\mathbb{Z}_2$ center of $SO(N_f)$. 
    
    \item For $N_f=0$ mod 4, there is no relations between the backgrounds, and the time-reversal symmetry is $T^2=1$.
\end{itemize}

\subsubsection{Anomaly for $T^2={\cal M}$: $N_f=2$ mod 4}

Since $w_1^2=w_2^f=0$ is trivial, the anomaly reduces to
\begin{align}\label{eqn:fullanomso2mod4}
    &
    \frac{N}{2}\pi\int w_2(SO(N_f))^2
    +\frac{N}{2}\pi\int w_3\cup B
    \cr 
&=   
\frac{N}{2}\pi\int w_2(SO(N_f))\cup (w_2+w_1^2)+\frac{N}{2}\pi\int w_3\cup B
    ~.
\end{align}
 This corresponds to the class $\ell=0$ of the classification of bosonic bulk 3+1d SPT phases discussed in \cite{Cordova:2023bja}. In addition, the flavor symmetry has a mixed anomaly with time-reversal symmetry when $N=2$ mod 4.

\subsubsection{Anomaly for $T^2=+1$: $N_f=0$ mod 4}

In this case, both $w_1^2$ and $w_2^f$ are nontrivial and independent of each other. The 
 anomaly is
\begin{align}\label{eqn:fullanomso0mod4}
    &
    \frac{2\pi}{4}\frac{NN_f}{8}\int {\cal P}(w_2)+
    \frac{NN_f}{8}\pi\int \left((w_2^f)^2+w_2^f\cup w_2\right)+\frac{N}{2}\pi\int w_2(SO(N_f))^2\cr
    &\quad +\frac{N}{2}\pi\int \left(w_3^f+w_3\right)\cup B
    -\frac{2\pi}{4}\frac{NN_f}{8}\int (p_1/3)\cr 
    &=\frac{NN_f}{8}\pi\int w_1^4+\frac{NN_f}{8}\pi\int w_2^f\cup w_1^2\cr
    &\quad +\frac{N}{2}\pi\int \left(w_3^f+w_3\right)\cup B+\frac{N}{2}\pi\int w_2(SO(N_f))\cup (w_2+w_1^2)~.
\end{align}
This represents $c_-=0$ and $\nu=NN_f\in 8\mathbb{Z}$.
In addition, the flavor symmetry has a mixed anomaly with the time-reversal symmetry when $N=2$ mod 4.

 We note that when $NN_f=0$ mod 16, $\nu=0$. On the other hand, there can still be anomalies involving the flavor symmetry when $N=2$ mod 4. Thus the time-reversal and flavor symmetry is an obstruction to symmetric mass generation without breaking the flavor symmetry even when the system has a multiple of 16 Majorana fermions.

\section{QED$_3$ with $N_f$ fermions}
\label{sec:U(1)}

Let us consider time-reversal symmetric $U(1)$ gauge theory with $N_f$ fermions. 
The theory is bosonic: the fermion parity $(-1)^F$ is part of the gauge group, all gauge invariant local operators are bosons, including the monopole operators. 
This corresponds to the case $N=2$ in section \ref{sec:SO}. For the theory to have time-reversal symmetry $N_f$ needs to be even.

\subsection{Symmetry}

When $N_f=2$ mod 4, the time-reversal symmetry satisfies $T^2=(-1)^m$ for $U(1)$ magnetic charge $m$, while $T^2=1$ for $N_f=0$ mod 4 \cite{Witten:2016cio,Cordova:2017kue}.
The theory has $U(N_f)/\mathbb{Z}_{N_f/2}$ unitary symmetry \cite{Benini:2017dus,Cordova:2017kue}. 
The symmetry structure is
\begin{equation}\label{eqn:QED3symmetry}
    {U(1)\times U(N_f)\times \widetilde{\text{Lorentz}}\over \mathbb{Z}_{N_f/2}\times \mathbb{Z}_2}~,
\end{equation}
where $\widetilde{\text{Lorentz}}$ is an extension of the Lorentz group depending on how time-reversal fractionalize on the charged fermions. We will take the fractionalization $T^2=(-1)^{mN_f/2}(-1)^F$.

The global symmetry is
\begin{equation}
    {U(N_f)\times \widetilde{\text{Lorentz}}\over \mathbb{Z}_{N_f/2}\times \mathbb{Z}_2}~.
\end{equation}
There are two discrete fluxes $w_2^f,w_2$ for the $\mathbb{Z}_{N_f/2}\times \mathbb{Z}_2$ quotient.
The quantization of the dynamical gauge field is modified to be
\begin{equation}
    \frac{da}{2\pi}=\frac{1}{N_f}\left(2w_2^f+ \frac{N_f}{2}  w_2\right)\text{ mod }1~.
\end{equation}

\subsection{Anomaly}

For the theory to have time-reversal symmetry, we need to add the regulating term
\begin{equation}
    U(1)_{N_f/2}\times U(N_f)_{1/2}+N_f\text{CS}_\text{grav}~.
\end{equation}
Such term is not well-defined, but it depends on the bulk as
\begin{align}
    &\frac{N_f\pi}{2}\int \frac{da}{2\pi}\frac{da}{2\pi}+
    I_{\theta_f=\pi}+N_f\frac{\pi}{8}\int (p_1/3)\cr 
    &=\frac{N_f}{2}\pi\int 
    \left(\frac{da}{2\pi}-\frac{1}{N_f}\left(2w_2^f+\frac{N_f}{2}w_2\right)\right) ^2
    +\frac{N_f}{2}\pi\int \left(\frac{da}{2\pi}-\frac{1}{N_f}\left(2w_2^f+\frac{N_f}{2}w_2\right)\right)w_2\cr 
    &\quad +\frac{N_f}{2}\frac{\pi}{4}\int {\cal P}(w_2)-\frac{2\pi}{N_f}\int {\cal P}(w_2^f)+I_{\theta_f=\pi}-\frac{N_f\pi}{8}\int (p_1/3)~.
\end{align}
To cancel the dependence on the dynamical field $a$, we introduce the background $B$ for the magnetic symmetry, which couples to the theory by the topological term
\begin{equation}
    \frac{1}{2\pi}\int_{3d} da B~.
\end{equation}
To cancel the $a$ dependence in the bulk term, the background $B$ satisfies
\begin{equation}
    dB=\frac{N_f}{2}\pi w_1^2\text{ mod }2\pi~.
\end{equation}
The background implies that $T^2=1$ for $N_f=0$ mod 4, but $T^2=(-1)^m$ for $N_f=2$ mod 4, where $m$ is the magnetic charge.
Since $da/2\pi$ is fractional, the topological term with $B$ also contributes to additional anomaly:
\begin{equation}
    S_\text{mag}=\int \left(\frac{2dw_2^f}{N_f}+\frac{dw_2}{2}\right)B~,
\end{equation}
where the terms in the bracket are integer cocycles.
The total anomaly is\footnote{
The action $\frac{\pi}{2}\int {\cal P}(x_2)$ for $\mathbb{Z}_2$ 2-cocycle $x_2$ can be defined on unorientable 4-manifolds using the quadratic function \cite{Browder:1969} (see \cite{Hsin:2021qiy} for related applications in gauge theories).
}
\begin{equation}\label{eqn:fullanomU(1)}
    \frac{N_f}{2}\frac{\pi}{2}\int {\cal P}(w_1^2)+I_{\theta_f=\pi}-\frac{2\pi}{N_f}\int {\cal P}(w_2^f)+\int \left(\frac{2dw_2^f}{N_f}+\frac{dw_2}{2}\right)B~.
\end{equation}
\begin{itemize}
    \item When $N_f=2$ mod 4, time-reversal symmetry satisfies $T^2=(-1)^m$, and $w_1^2$ is trivial. 
    There is anomaly for flavor symmetry and magnetic symmetry.
    
    \item When $N_f=0$ mod 4, time-reversal symmetry satisfies $T^2=1$, and $w_1^2$ is non-trivial. The time-reversal anomaly corresponds to 
    $\nu=2N_f\in 8\mathbb{Z}$. In addition, there is anomaly for flavor symmetry and magnetic symmetry.

\end{itemize}

\subsection{Anomaly of different anti-unitary symmetries $T'=TU$}

We can combine the time-reversal symmetry $T$ that commutes with the gauge group with another unitary symmetry $U$ to define a new time-reversal symmetry $T'=TU$. The anomaly of $T'$ can be derived from that of $T,U$. Let us illustrate this procedure using the example of QED$_3$ with $N_f=4$ Dirac fermions.
The theory has $T^2=1$ time-reversal symmetry, and is proposed to describe Dirac spin liquid in certain Heisenberg antiferromagnet (e.g. \cite{PhysRevX.10.011033}).\footnote{
We thank Meng Cheng for bringing up this example.
}
The anomalies of $N_f=4$ has been computed in \cite{Calvera:2021jqi,Zou:2021dwv}, where the computation considers the background gauge field of internal symmetry to be $SU(4)\times U(1)$ without $\mathbb{Z}_2$ quotient, and the $\mathbb{Z}_2$ charge conjugation.
We remark that the theory is also used to describe an ``unnecessary" quantum critical point on the lattice \cite{Zhang:2024hiv}.

\subsubsection{New $T'$ that combines charge conjugation symmetry}

If we change the time-reversal symmetry with additional charge conjugation ${\cal C}$, then the free fermion does not contribute to anomaly $\nu=0$.\footnote{
The anomaly of time-reversal symmetry ${\cal T}^2=(-1)^F$ is computed by $\nu=n_+-n_-$ mod 16, where $n_\pm$ are the number of massless Majorana fermions transformed by ${\cal T}$ with a positive and negative sign, respectively (see e.g. \cite{Witten:2016cio}).
} This means that there is a mixed anomaly between time-reversal symmetry and the charge conjugation symmetry.
Let us compute the anomaly for these $\mathbb{Z}_2$ symmetries. We will omit the flavor symmetry.

In the presence of the charge conjugation background $B^{\cal C}$, the $U(1)$ bundle is twisted to be $O(2)$ bundle with $w_1(O(2))=B^{\cal C}$.
The symmetry structure is
\begin{equation}
    {O(2)\times \widetilde{\text{Lorentz}}\over \mathbb{Z}_2}~.
\end{equation}
We note that the center of $O(2)$ is $\mathbb{Z}_2$, which is identified with $(-1)^F$.

To preserve time-reversal symmetry in $O(2)$ gauge theory with $N_f=4$ massless fermions, we need to add the counterterm term
\begin{equation}
    O(2)_{2,2}\times 4\text{CS}_\text{grav}~,
\end{equation}
where the Chern-Simons term $O(2)_{2,2}$ includes both the Chern-Simons term for $SO(2)$ and Chern-Simons term for $\mathbb{Z}_2$ gauge field, and we use the same notation as \cite{Cordova:2017vab}.
The above counterterm is not well-defined, but it depends on the bulk as
\begin{align}\label{eqn:anomCandT}
    &\frac{2\pi}{4}\int {\cal P}(w_2)+
    \pi\int B^{\cal C}\cup \text{Bock}(w_2)+\pi\int (B^{\cal C})^4-\frac{\pi}{2}\int (p_1/3)\cr 
    &=
    \pi\int \left((B^{\cal C})^4+w_1^4\right)+\pi\int (B^{\cal C})^2\cup w_2~,
\end{align}
where in the first line $\pi\int (B^{\cal C})^2=\pi\int (dB^{\cal C}/2)\cup (dB^{\cal C}/2)$ comes from the $\mathbb{Z}_2$ Chern-Simons term in $O(2)_{2,2}$. The $\frac{2\pi}{4}\int {\cal P}(w_2)+\pi\int B^{\cal C}\cup \text{Bock}(w_2)$ comes from the property that $O(2)_{2,0}$ has $\mathbb{Z}_4$ one-form symmetry \cite{Cordova:2017vab} which is the extension of the center $\mathbb{Z}_2$ symmetry and quantum $\mathbb{Z}_2$ symmetry generated by $\oint B^{\cal C}$. Thus the
anomaly from $O(2)_{2,0}$ is
\begin{equation}
  \frac{2\pi}{4}\int {\cal P}(w_2)+\pi\int B^{\cal C}\cup \text{Bock}(w_2)
 = \frac{2\pi}{4}\int {\cal P}(w_2)+\pi\int B^{\cal C}\cup \left(w_3+w_1w_2\right)
  ~,
\end{equation}
where we note that the first term has order 4 and is the ordinary Pontryagin square of $w_2$, 
and the last term will force $\text{Bock}(w_2)$ to be exact if we gauge the charge conjugation symmetry with dynamical field. This means that the $(-1)$ symmetry that flips the sign of the fermion, which is in the center of the gauge group, will not give a $\mathbb{Z}_2$ center one-form symmetry of the pure $O(2)_{2,2}$ Chern-Simons term after gauging the charge conjugation symmetry, but instead it becomes a $\mathbb{Z}_4$ symmetry.

\paragraph{Time-reversal anomaly for $T'=T{\cal C}$}

The anomaly for $T'=T{\cal C}$ is given by setting $B^{\cal C}=w_1$ in (\ref{eqn:anomCandT}), which gives trivial anomaly
using $w_1w_2=0$ mod 2 on four-manifolds. The absence of time-reversal anomaly agrees with the free fermion computation.

The discussion can be generalized straightforwardly to other even $N_f$. 
Since the time-reversal combines charge conjugation symmetry that does not commute with the gauge group, the time-reversal symmetry $T'$ has order 2 for any even $N_f$, since $\pi\int w_1^2\frac{da}{2\pi}$ is trivial for $U(1)\rtimes \mathbb{Z}_2^T$ (see e.g. \cite{Kapustin:2014gma}). 
 This can be seen alternatively by studying the action of time-reversal symmetry on the states in unit monopole background, described by the Fock space of fermion zero modes $\alpha_I$ with $I=1,2,\cdots , N_f$ \cite{Cordova:2017kue}.
The time-reversal symmetry $T$ that commutes with the gauge group (i.e. the electric charge flips sign under $T$) acts on the zero modes as $T\alpha_IT^{-1}=\alpha_I^\dag$, while the time-reversal symmetry $T'=T{\cal C}$ leaves the zero modes invariant. Thus $T'$ leaves the Fock space vacuum $|0\rangle$ invariant. On the other hand, $T$ maps $|0\rangle$ which is annihilated by all $\alpha_I$, to the top state $\alpha_1^\dag\alpha_2^\dag\cdots \alpha_{N_f}^\dag|0\rangle$ annihilated by all $\alpha_I^\dag$. 
The monopoles correspond to the state obtained from $|0\rangle$ by $N_f/2$ creation operators. From the symmetry action one can derive $T^2=(-1)^{N_f/2}$ and $T'^2=+1$ on the monopole state \cite{Cordova:2017kue}.

\subsubsection{New $T'$ that combines flavor charge conjugation symmetry}

Consider the flavor charge conjugation ${\cal C}_f$ that only flips the sign of the first flavor of Dirac fermion. The symmetry is not in $SU(N_f)$, but in $SU(N_f)\rtimes \mathbb{Z}_2$. We will call it the flavor charge conjugation symmetry.

\paragraph{Deformed symmetry on monopoles}

In general $N_f$ flavor QED$_3$, if we combine $T$ with ${\cal C}_f$ to define new time-reversal symmetry $T'=T{\cal C}_f$, the new time-reversal symmetry $T'$ satisfies $T'^2=(-1)^{N_f/2+1}(-1)^m$ \cite{Cordova:2017kue}. Let us provide another argument here.

In the presence of the background $\mathbb{Z}_2$ gauge field $A$ for ${\cal C}_f$, one flavor of Dirac fermion couples to $a+\pi A$ for the dynamical gauge field $a$, while other fermions couple to $a$. Thus to preserve time-reversal symmetry, we need to add the counterterm
\begin{equation}
    U(1)_{N_f/2}^{a}+\frac{\pi}{8}AdA+\frac{1}{4}Ada+N_f\text{CS}\text{grav}~.
\end{equation}
Such counterterm is not well-defined, but it depends on the bulk. Let us consider pin$^+$ manifolds for simplicity. Then the bulk dependence is
\begin{equation}
    \frac{N_f}{2}\pi\int \frac{da}{2\pi}w_1^2+\frac{\pi}{2}\int {\cal P}(dA/2)+\pi\int w_1A\frac{da}{2\pi}-\frac{N_f}{2}\frac{\pi}{2}\int {\cal P}(w_1^2)~.
\end{equation}
For the dynamical field $da$ to be independent of the bulk, $\frac{N_f}{2}w_1^2+w_1A$ is trivial.
In other words, the symmetries are extended, with extension specified by the two-cycle $\frac{N_f}{2}w_1^2+w_1A$.
In particular, the $w_1A$ cocycle implies that $T$ and ${\cal C}_f$ does not commute in the presence of monopole, which is consistent with \cite{Cordova:2017kue}.

\paragraph{Time-reversal anomalies}

If we put the theory on unorientable manifolds using the new time-reversal symmetry $T'=T{\cal C}_f$, the background is $A=w_1$. This means that $\frac{N_f+2}{2}w_1^2$ is trivial. In other words, $T'^2=(-1)^{(N_f/2+1)m}$. The anomaly is 
\begin{equation}
    \frac{2-N_f}{2}\frac{\pi}{2}\int {\cal P}(w_1^2)~,
\end{equation}
which corresponds to $\nu=-(2N_f-4)$. This is the same result from the free fermion computation, where the number of Majorana fermions that do or do not flip sign under time-reversal is $2$ and $2N_f-2$, and the difference is $\nu=2-(2N_f-2)=4-2N_f\in 8\mathbb{Z}$ for $N_f=2$ mod 4 where $T'^2=1$.

\paragraph{New $T'$ that combines both charge conjugation symmetries}

In \cite{PhysRevX.10.011033} the magnetic charge is odd under time-reversal, while here the magnetic charge is even under time-reversal.  
In addition, the time-reversal symmetry flips the sign of 3 out of the 6 monopoles that form 2-index antisymmetric representation of $SU(4)$ flavor symmetry. The 6 monopoles carry $SU(4)$ fundamental indices ${\cal M}_{1,2},{\cal M}_{1,3},{\cal M}_{1,3},{\cal M}_{2,3},{\cal M}_{2,4},{\cal M}_{3,4}$. Thus the internal symmetry is the flavor charge conjugation symmetry ${\cal C}_f$ that flips the sign of index $1$.
In other words, the time-reversal symmetry is $T'=T{\cal C}{\cal C}_f$. By combining the previous two cases, one finds both $w_2^2$ and $w_1^4$ anomalies are absent, in agreement with \cite{PhysRevX.10.011033,Calvera:2021jqi,Zou:2021dwv}.

\section{Examples with $c_-\neq 0$ mod 8}
\label{sec:ranktwo}

In all of the previous examples with fundamental fermions there is no $w_2^2$ anomaly, i.e. $c_-=0$ mod 8. 
We will prove that this is the case in any bosonic theories with even number of charged Dirac fermions and without one-form symmetry.
We will also give examples with nonzero $w_2^2$ anomaly. In particular, we will show there are bosonic gauge theories with charged fermions that have a multiple of 16 Majorana fermion with $\nu=0$ but nonzero $w_2^2$ anomaly $\alpha=1$.

\subsection{$SU(4k)$ QCD with symmetric tensor fermion}

\subsubsection{Example: $SU(4)$}

Consider $SU(4)$ gauge theory with a massless Dirac fermion in 2-index symmetric tensor representation.
The dynamics of the theory is studied in \cite{Choi:2018tuh}.
The Dynkin index of the representation is $(4+2)/2=3$, and there are in total $10$ Dirac fermions.  
The $\mathbb{Z}_2$ subgroup in the center of the gauge group acts trivially on the matter, and the theory has $\mathbb{Z}_2$ one-form symmetry.
The theory is bosonic: the fermion parity is identified with a $\mathbb{Z}_4$ rotation in the center of the gauge group.
This implies that the theory has two-group symmetry. The background $B_2$ for the one-form symmetry satisfies
\begin{equation}
    dB_2=\text{Bock}(w_2)=w_3+w_1w_2~.
\end{equation}
The time-reversal symmetry on gauge invariant local operators satisfy $T^2=+1$.
We note that it is still meaningful to ask if the theory has the bosonic $w_2^2$ anomaly for the two-group symmetry.

The symmetry structure is
\begin{equation}\label{eqn:SU4symmetry}
    {SU(4)/\mathbb{Z}_2\times \widetilde{\text{Lorentz}}\over \mathbb{Z}_2}~,
\end{equation}
where we take the fractionalization of time-reversal symmetry on the charged fermion to be $T^2=(-1)^F$. 

For the theory to be time-reversal symmetric, we need to add the following counterterm
\begin{equation}
    SU(4)_3\times 10\text{CS}_\text{grav}~.
\end{equation}
For the twisted bundle in (\ref{eqn:SU4symmetry}),
such counterterm is not well-defined and depends on the bulk by\footnote{
In particular, the nontrivial $w_2^2$ anomaly in the equation can be verified on $\mathbb{CP}^2$ where $w_2=c_1$ mod 2.
}
\begin{equation}
    2\pi\cdot 3\cdot \frac{4-1}{2\cdot 4}\int {\cal P}(2B_2+w_2)-10\cdot \frac{\pi}{8}\int (p_1/3)
    =\pi\int \left(B_2\cup w_1^2+w_2^2\right)+\frac{\pi}{2}\int {\cal P}(w_1^2)~.
\end{equation}
 We note that the anomaly is invariant under shifting $B_2\rightarrow B_2+w_2$. 

It would be interesting to constrain the infrared dynamics.The presence of the $w_2^2$ anomaly imposes nontrivial constraint on the infrared dynamics of the theory. For instance, it was proposed in \cite{Choi:2018tuh} that at low energy the theory flows to the bosonic theory $U(3)_{3,0}={SU(3)_3\times S^1\over \mathbb{Z}_3}$ that describes the TQFT $SU(3)_3/\mathbb{Z}_3$ coupled to Goldstone boson. The Chern-Simons term alone only contributes $c_-=\pm 2$ mod 8.\footnote{
In fact, the TQFT alone is not time-reversal symmetric as a bosonic theory \cite{Hsin:2016blu}.
}

\subsubsection{Generalization}

The discussion can be generalized to $SU(4k)$ gauge theory with a 2-index symmetric tensor Dirac fermion. 
The Dynkin index is $(4k+2)/2=2k+1$, and there are in total $2k(4k+1)$ Dirac fermions.
To preserve time-reversal symmetry, we need to add counterterm
\begin{equation}
    SU(4k)_{2k+1}\times 2k(4k+1)\text{CS}_\text{grav}~.
\end{equation}
Such counterterm is not well-defined, but depends on the bulk as
\begin{align}
    &2\pi(2k+1)\frac{4k-1}{8k}k^2\int {\cal P}(2B+w_2)
    -
    2k(4k+1)\frac{\pi}{8}\int (p_1/3)\cr
    &=k\pi\int B\cup w_1^2+\frac{k(k+1)}{2}\pi\int w_2^2-k\frac{\pi}{2}\int {\cal P}(w_1^2)~.
\end{align}
For $k=4n+2$, i.e. $SU(16n+8)$ gauge theory, the anomaly is $\alpha=1,\beta=1$:
\begin{equation}
    \pi\int w_2^2+\pi\int w_1^4~.
\end{equation}

In particular, we can stack $SU(16n+8)$ gauge theory with tensor fermion with additional bosonic QCD with vector fermions where the one-form symmetry does not act, e.g. in section \ref{sec:SU} and section \ref{sec:Sp} that have anomaly $\alpha=0,\beta=1$, i.e. $c_-=4$ mod 8.
An example is $SU(2)$ gauge theory with two fundamental fermions.
Then the total theory has a multiple of 16 Majorana fermions with $\nu=0$, but the time-reversal anomaly $w_2^2$ is nontrivial.
We note that the value of $c_-$ is different from the free fermion contribution: with $c_-^\text{free}=1/2$ for each massless Dirac fermion, and take $n=0$ and stacked with $SU(2)$ QCD with two fundamental fermions, this would give $\frac{1}{2}\cdot (36+4)=40=0$ mod 8, which is different from the correct anomaly $c_-=4$ mod 8.

We note that the theory with massless symmetric tensor fermion is proposed in \cite{Choi:2018tuh} to flow to the bosonic theory $U(2k+1)_{2k+1,0}={SU(2k+1)_{2k+1}\times S^1\over \mathbb{Z}_{2k+1}}$ that describes the TQFT $SU(2k+1)_{2k+1}/\mathbb{Z}_{2k+1}$ coupled to Goldstone bosons.
The TQFT has $c_-=\pm 2k$ mod 8. For $k=4n+2$, it is $c_-=\pm 4$ mod 8. The TQFT for $SU(16n+8)$ gauge theory is time-reversal symmetric as a bosonic TQFT by the level/rank duality of bosonic Chern-Simons theories \cite{Aharony:2016jvv,Hsin:2019gvb}, and the $w_2^2$ anomaly matches with the anomaly in the UV gauge theory.

\subsection{$SU(4k)$ QCD with antisymmetric tensor fermion}

\subsubsection{Example: $SU(8)$}
Consider $SU(8)$ gauge theory with a massless Dirac fermion in the 2-index antisymemtric tensor representation. 
The dynamics of the theory is studied in \cite{Choi:2018tuh}.
The Dynkin index is $(8-2)/2=3$, and there are in total $8\times 7/2=28$ Dirac fermions.
The $\mathbb{Z}_2$ subgroup in the center of the gauge group acts trivially on the matter, and the theory has $\mathbb{Z}_2$ one-form symmetry.
The theory is bosonic: the fermion parity is identified with a $\mathbb{Z}_4$ rotation in the center of the gauge group.
This implies that the theory has two-group symmetry. The background for the $\mathbb{Z}_2$ one-form symmetry satisfies
\begin{equation}
    dB_2=\text{Bock}(w_2)=w_3+w_1w_2~.
\end{equation}
The time-reversal symmetry on gauge invariant local operators satisfy $T^2=+1$.
We note that it is still meaningful to ask if the theory has the bosonic $w_2^2$ anomaly for the two-group symmetry.

The symmetry structure is
\begin{equation}\label{eqn:SU8symmetry}
    {SU(8)/\mathbb{Z}_2\times \widetilde{\text{Lorentz}}\over \mathbb{Z}_2}~,
\end{equation}
where we take the fractionalization of time-reversal symmetry on the charged fermion to be $T^2=(-1)^F$. 

For the theory to be time-reversal symmetric, we need to add the following counterterm
\begin{equation}
    SU(8)_3\times (28\text{CS}_\text{grav})~.
\end{equation}
For the twisted bundle in (\ref{eqn:SU8symmetry}),
such counterterm is not well-defined and depends on the bulk by
\begin{equation}
    2\pi\cdot 3\cdot \frac{8-1}{2\cdot 8}\cdot (8/4)^2\int {\cal P}(2B_2+w_2)-28\cdot \frac{\pi}{8}\int (p_1/3)
    =\pi\int w_2^2+\pi\int w_1^4 ~.
\end{equation}
This corresponds to the time-reversal anomaly $\alpha=1,\beta=1$.

It would be interesting to constrain the infrared dynamics.The presence of the $w_2^2$ anomaly imposes nontrivial constraint on the infrared dynamics of the theory. For instance, it was proposed in \cite{Choi:2018tuh} that at low energy the theory flows again to the bosonic theory $U(3)_{3,0}={SU(3)_3\times S^1\over \mathbb{Z}_3}$ that describes the TQFT $SU(3)_3/\mathbb{Z}_3$ coupled to Goldstone boson. The Chern-Simons TQFT only contributes $c_-=\pm 2$, and it is not time-reversal symmetric as a bosonic TQFT.

\subsubsection{Generalization}

The discussion can be generalized to $SU(4k)$ gauge theory with a 2-index antisymmetric tensor Dirac fermion. 
The Dynkin index is $(4k-2)/2=2k-1$, and there are in total $2k(4k-1)$ Dirac fermions.
To preserve time-reversal symmetry, we need to add counterterm
\begin{equation}
    SU(4k)_{2k-1}\times 2k(4k-1)\text{CS}_\text{grav}~.
\end{equation}
Such counterterm is not well-defined, but depends on the bulk as
\begin{align}
    &2\pi (2k-1)\frac{4k-1}{8k}k^2\int {\cal P}(2B_2+w_2)
    -2k(4k-1)\frac{\pi}{8}\int (p_1/3)\cr 
    &=k\pi \int B_2\cup w_1^2+\frac{k(k-1)}{2}\pi\int w_2^2
    -k\frac{\pi}{2}\int {\cal P}(w_1^2)~.
\end{align}
For $k=4n+2$, i.e. $SU(16n+8)$ gauge theory, the anomaly becomes $\alpha=1,\beta=1$:
\begin{equation}
    \pi\int w_2^2+\pi\int w_1^4~.
\end{equation}

In particular, we can stack $SU(16n+8)$ gauge theory with tensor fermion with additional bosonic QCD with vector fermions where the one-form symmetry does not act, e.g. in section \ref{sec:SU} and section \ref{sec:Sp} that have anomaly $\alpha=0,\beta=1$, i.e. $c_-=4$ mod 8.
An example is $SU(2)$ gauge theory with two fundamental fermions.
Then the total theory has a multiple of 16 Majorana fermions with $\nu=0$, but the time-reversal anomaly $w_2^2$ is nontrivial.
We note that the value of $c_-$ is different from the free fermion contribution: with $c_-^\text{free}=1/2$ for each massless Dirac fermion, and take $n=0$ and stacked with $SU(2)$ QCD with two fundamental fermions, this would give $\frac{1}{2}\cdot (28+4)=16=0$ mod 8, which is different from the correct anomaly $c_-=4$ mod 8.

We note that the theory with massless antisymemtric tensor fermion is proposed in \cite{Choi:2018tuh} to flow to the bosonic theory $U(2k-1)_{2k-1,0}={SU(2k-1)_{2k-1}\times S^1\over \mathbb{Z}_{2k-1}}$ that describes the TQFT $SU(2k-1)_{2k-1}/\mathbb{Z}_{2k-1}$ coupled to Goldstone bosons.
The TQFT has $c_-=\pm(2k-2)$ mod 8. For $k=4n+2$, it is $c_-=\pm 2$ mod 8. The TQFT is not time-reversal symmetric as a bosonic TQFT.

\subsection{Discrete gauge theory with fermions}

Here we provide a universal explanation for the absence of $w_2^2$ anomaly in gauge theory without one-form symmetry in section \ref{sec:SU}-\ref{sec:SO}. For simplicity, we will assume the total number of Dirac fermions is even,
which is the case in all our examples in section \ref{sec:SU}-\ref{sec:SO}.

To compute the anomalies, we can introduce adjoint scalars $\phi^A$ that are even under time-reversal symmetry.\footnote{
This is the case for the time-reversal symmetry $T$ that commutes with the gauge group. For instance, the spatial components of the gauge field, which is also in the adjoint representation, transform as $a_i(t,\vec{x})\rightarrow a_i(-t,\vec{x})$.
} The theory is still bosonic, since the fermion parity, which is in the center of the gauge group, does not act on the adjoint scalars. We do not turn on any Yukawa couplings. When the scalars do not condense, the theory with scalars flow to the original gauge theories with charged fermions. When the scalars condense, we can use sufficiently number of scalars and generic potential such that the gauge group is broken to the center. Then the anomaly can be computed in Abelian gauge theories coupled to fermions. The Abelian gauge groups can be product of $U(1)$ or $\mathbb{Z}_{2k}\supset \mathbb{Z}_2^f$ fermion parity.
The former is already discussed in section \ref{sec:U(1)} where it is shown that the $w_2^2$ anomaly is absent. Here we will focus on the latter case. 
Next, We further add charge-two Higgs scalars $\Phi$ with real condensate and no Yukawa couplings such that the gauge group is further broken down to $\mathbb{Z}_2^f$ fermion parity, while preserving the original time-reversal symmetry $\Phi(t,\vec{x})\rightarrow \Phi^*(-t,\vec{x})$.

The microscopic theory without one-form symmetry flows to
$\mathbb{Z}_2^f$ gauge theory coupled to $N_f$ Dirac fermions with even $N_f$.
To preserve time-reversal symmetry, we need to add the counterterm
\begin{equation}
    (\mathbb{Z}_2^f)_{N_f}+N_f\text{CS}_\text{grav}~.
\end{equation}
This is not a well-defined counterterm, but depends on the bulk s
\begin{align}
    &\frac{N_f}{2}\pi\int \frac{da}{2\pi}\frac{da}{2\pi}-N_f\frac{\pi}{8}\int (p_1/3) \cr 
    &=
        \frac{N_f}{2}\pi\int \left(\frac{da}{2\pi}-\frac{1}{2}w_2\right)^2+\frac{N_f}{2}\pi\int \left(\frac{da}{2\pi}-\frac{1}{2}w_2\right) w_2
        +\frac{N_f}{2}\frac{\pi}{2}\int {\cal P}(w_1^2)\cr 
        &=
        \frac{N_f}{2}\pi\int \left(\frac{da}{2\pi}-\frac{1}{2}w_2\right)w_1^2
        +\frac{N_f}{2}\frac{\pi}{2}\int {\cal P}(w_1^2)~,
\end{align}
where we embed the $\mathbb{Z}_2$ gauge field in a $U(1)$ gauge field $a$ with holonomy $0,\pi$ mod $2\pi$.
In terms of the $\mathbb{Z}_2$ gauge field $\bar a\sim a/\pi=0,1$ mod 2, the first term in the last line is
\begin{equation}
\frac{N_f}{2}    \pi\int \bar a^2w_1^2=
\frac{N_f}{2}
\pi\int Sq^1(\bar aw_1^2)=\pi\int \bar a w_1^3~,
\end{equation}
where we used the Wu formula $w_1\cup x_3=Sq^1 x_3$ on four-manifolds, and $Sq^n(x\cup y)=\sum_{i=0}^n Sq^i(x)\cup Sq^{n-i}(y)$. If $da$ comes from magnetic flux in the microscopic gauge theory, this implies that for $N_f=2$ mod 4, $T^2={\cal M}$ for $\mathbb{Z}_2$ (subgroup) magnetic symmetry ${\cal M}$. For $N_f=0$ mod 4 the time-reversal symmetry is still $T^2=1$ in the UV gauge theory. In both cases the $w_2^2$ anomaly is absent. This is consistent with the results in section \ref{sec:SU}-\ref{sec:U(1)}.

\section*{Acknowledgement}

We thank Ryan Thorngren, Cenke Xu and Yi-Zhuang You for related discussions. We thank Meng Cheng and Chong Wang for discussions and comments on a draft. 
We thank Thomas Dumitrescu, Jaume Gomis and Ho tat Lam for comments on a draft.
The work was supported by Simons Collaboration of Global Categorical Symmetry, Department of Mathematics King's College London, and
also supported in part by grant NSF PHY-2309135 to the Kavli Institute for Theoretical Physics (KITP).
We thank Kavli Institute for Theoretical Physics for hosting the program ``Correlated Gapless Quantum Matter'' in 2024, during which part of the work is completed.

\appendix

\section{Useful Mathematical Identities for Characteristic Classes}
\label{sec:mathidentity}

Here we summarize some useful mathematical relation for Stiefel-Whitney classes and Pontryagin classes.
\begin{itemize}
    \item Wu formula. On a general closed $D$-manifold and $\mathbb{Z}_2$ $n$-cocycle $x_n$, $Sq^{D-n}x_n=x_n\cup_{2n-D} x_n=v_{D-n}\cup x_n$ mod 2, where $v_{D-n}$ is the $(D-n)$th Wu class (see e.g. \cite{milnor1974characteristic}). For $D=4$ and $n=2$, this is $x_2\cup x_2=Sq^2(x_2)=(w_2+w_1^2)\cup x_2$ mod 2, where we used the second Wu class $v_2=w_2+w_1^2$.

\item Relation between Stiefel-Whitney classes. 
Using $Sq^m x_k=0$ for $m>k$ and the Wu formula, one finds $Sq^{D-k}x_k=0=v_{D-k}\cup x_k$ for $D>2k$, Thus $v_{m}=0$ for $m>D/2$. For instance when $D=4$, this gives $v_3=0=w_1w_2$. Combining this with $v_4=0$ gives $w_4+w_2^2+w_1^4=0$. See e.g. \cite{Kapustin:2014tfa,Wen:2014zga} for a summary of the relations between Stiefel-Whitney classes in different dimensions.

\item Relation between Pontryagin classes and Stiefel-Whitney classes for general $BO$ cohomology \cite{Wu:1954}
\begin{align}
    &{\cal P}(w_{2k+1})=\text{Bock}\left(Sq^{2k}w_{2k+1}\right)+2w_1Sq^{2k}w_{2k+1}\quad \text{ mod }4~,\cr 
    &{\cal P}(w_{2k})=p_k+\text{Bock}\left(w_{2k-1}w_{2k}\right)+2w_1Sq^{2k-1}w_{2k}+2\sum_{j=0}^{k-1}w_{2j}w_{4k-2j}\quad \text{ mod }4~,
\end{align}
where ${\cal P}$ is the Pontryagin square operation,
Bock is the Bockstein homomorphism for $2\mathbb{Z}\rightarrow\mathbb{Z}\rightarrow \mathbb{Z}_2=\mathbb{Z}/2\mathbb{Z}$.
In particular, for the tangent bundle in $D=4$ dimension, \begin{equation}
    {\cal P}(w_2)=p_1+2w_4\text{ mod }4~.
\end{equation}
Using $w_4+w_2^2+w_1^4=0$ mod 2 on four-manifolds, this becomes
\begin{equation}
    {\cal P}(w_2)=p_1+2w_2^2+2w_1^4\text{ mod }4
    \quad \Rightarrow\quad  p_1=3{\cal P}(w_2)+2w_1^4\text{ mod }4~.
\end{equation}

\end{itemize}

\bibliographystyle{utphys}
\bibliography{biblio}{}

\end{document}